\documentclass[aps, superscriptaddress ,twocolumn]{revtex4}
\usepackage{amsmath}
\usepackage{amsfonts}
\usepackage{graphicx}
\usepackage{hyperref}

\newcommand{\specialcell}[2][c]{%
  \begin{tabular}[#1]{@{}c@{}}#2\end{tabular}}

\begin{document}

\title{Tuning electronic correlations in transition metal
pnictides: chemistry beyond the valence count}

\author{E. Razzoli}
\affiliation{Swiss Light Source, Paul Scherrer Institute, CH-5232 Villigen PSI, Switzerland}
\affiliation{D{\'e}partement de Physique and Fribourg Center for Nanomaterials, Universit\'e de Fribourg, CH-1700 Fribourg, Switzerland}

\author{C. E. Matt}
\affiliation{Swiss Light Source, Paul Scherrer Institute, CH-5232 Villigen PSI, Switzerland}
\affiliation{Laboratory for Solid State Physics, ETH Z{\"u}rich, CH-8093 Z{\"u}rich, Switzerland}

\author{M. Kobayashi}
\affiliation{Swiss Light Source, Paul Scherrer Institute, CH-5232 Villigen PSI, Switzerland}
\affiliation{Department of Applied Chemistry, School of Engineering,
University of Tokyo, 7-3-1 Hongo, Bunkyo-ku, Tokyo 113-865 6, Japan}

\author{X.-P. Wang}
\affiliation{Beijing National Laboratory for Condensed Matter Physics, and Institute of Physics, Chinese Academy of Sciences, Beijing 100190, China}

\author{V. N. Strocov}
\affiliation{Swiss Light Source, Paul Scherrer Institute, CH-5232 Villigen PSI, Switzerland}

\author{A. van Roekeghem}
\affiliation{Centre de Physique Theorique, Ecole Polytechnique, CNRS, F-91128 Palaiseau Cedex, France}
\affiliation{Beijing National Laboratory for Condensed Matter Physics, and Institute of Physics, Chinese Academy of Sciences, Beijing 100190, China}

\author{S. Biermann}
\affiliation{Centre de Physique Theorique, Ecole Polytechnique, CNRS, F-91128 Palaiseau Cedex, France}
\affiliation{Coll\`ege de France, 11 place Marcelin Berthelot, 75005 Paris, France}
\affiliation{European Theoretical Synchrotron Facility, Europe}

\author{N. C. Plumb}
\affiliation{Swiss Light Source, Paul Scherrer Institute, CH-5232 Villigen PSI, Switzerland}

\author{M. Radovic}
\affiliation{Swiss Light Source, Paul Scherrer Institute, CH-5232 Villigen PSI, Switzerland}

\author{T. Schmitt}
\affiliation{Swiss Light Source, Paul Scherrer Institute, CH-5232 Villigen PSI, Switzerland}

\author{C. Capan}
\affiliation{Department of Physics and Astronomy, University of California Irvine, Irvine, California 92697, USA}

\author{Z. Fisk}
\affiliation{Department of Physics and Astronomy, University of California Irvine, Irvine, California 92697, USA}

\author{P. Richard}
\affiliation{Beijing National Laboratory for Condensed Matter Physics, and Institute of Physics, Chinese Academy of Sciences, Beijing 100190, China}
\affiliation{Collaborative Innovation Center of Quantum Matter, Beijing, China}

\author{H. Ding}
\affiliation{Beijing National Laboratory for Condensed Matter Physics, and Institute of Physics, Chinese Academy of Sciences, Beijing 100190, China}
\affiliation{Collaborative Innovation Center of Quantum Matter, Beijing, China}

\author{P. Aebi}
\affiliation{D{\'e}partement de Physique and Fribourg Center for Nanomaterials, Universit\'e de Fribourg, CH-1700 Fribourg, Switzerland}

\author{J. Mesot}
\affiliation{Swiss Light Source, Paul Scherrer Institute, CH-5232 Villigen PSI, Switzerland}
\affiliation{Laboratory for Solid State Physics, ETH Z{\"u}rich, CH-8093 Z{\"u}rich, Switzerland}
\affiliation{Institut de la Matiere Complexe, EPF Lausanne, CH-1015, Lausanne, Switzerland}

\author{M. Shi}
\affiliation{Swiss Light Source, Paul Scherrer Institute, CH-5232 Villigen PSI, Switzerland}

\date{\today}

\begin{abstract}
The effects of electron-electron correlations on the low-energy electronic structure and  their relationship with unconventional superconductivity are central aspects in the research on  the iron-based pnictide superconductors. 
Here we use soft X-ray angle-resolved photoemission spectroscopy (SX-ARPES) to  study how  electronic correlations evolve in different chemically substituted iron pnictides. We find that correlations are  intrinsically related to the effective filling of the correlated orbitals, rather than to the  filling obtained by valence counting. 
Combined density functional theory (DFT) and dynamical mean-field theory (DMFT) calculations capture these effects, reproducing the experimentally observed trend in the correlation strength. The   occupation-driven trend in the electronic correlation reported in our work supports the recently proposed connection between cuprate and pnictides phase diagrams.
\end{abstract} 

\maketitle

Many recent studies have focused on the importance of the occupancy of 
the bands close to the Fermi level ($E_F$)  in promoting the superconducting  state of the iron based superconducting pnictides \cite{DeMedici2014, Ye2014}. Most interestingly, when the filling is expressed with respect to the half-filled $3d^5$ bands, a striking similarity of the iron pnictide phase diagram to that of high-temperature superconducting cuprates has been pointed out \cite{Misawa2012,DeMedici2014}, 
suggesting links between strongly correlated Mott insulating behavior
and unconventional superconductivity.
In the iron pnictides, electronic Coulomb correlations are largely
driven by Hund's exchange coupling \cite{Haule2009, Aichhorn2010,
DeMedici2011} leading to a strong dependence even of the normal state properties on doping \cite{Werner2012} and possible orbital-selectivity in the strong coupling regime \cite{DeMedici2009}.
Theoretical models including the five-fold orbital manifold and
Hund's rule coupling in addition to local Hubbard interactions
\cite{Liebsch2010,
Lanata2013} 
reproduce the main consequences of the change of filling induced
by doping, in particular a regime with strongly enhanced effective 
masses when approaching the d$^5$ configuration, even at local
Hubbard interactions much smaller than the critical one.
However, while this picture works well in the case of substitution of Ba 
with K or of Fe with Co \cite{Xu2013},  it fails in case of isovalent substitution, e.g., for BaFe$_2$As$_{2-x}$P$_x$, where the correlations change with $x$ but the nominal filling of the correlated bands does not \cite{Ye2014}. 
From the theoretical side, it has also been pointed out that nominal
doping does not necessarily lead to a consistent change in band fillings
\cite{Wadati2010}.
Quite generally,  the link between the  correlation strength in the Fe-pnictides and the valence count remains elusive.


\begin{figure*}
\includegraphics[width=1\textwidth]{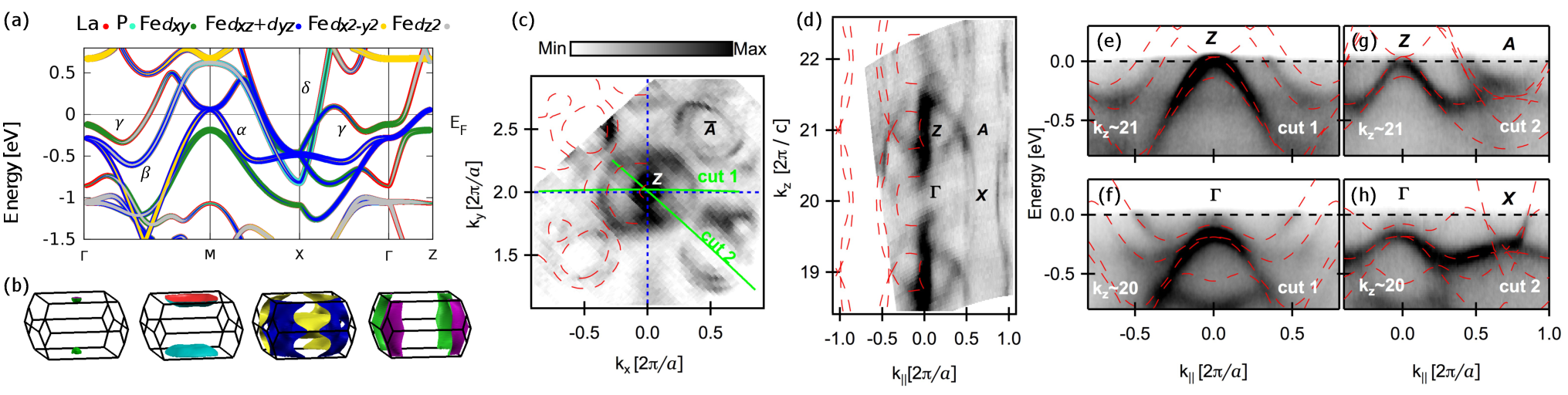}
\caption{ DFT band structure and SX-ARPES intensity maps of LaFe$_2$P$_2$. (a). Band structure along high symmetry lines.  The thickness of the coloured line is proportional to the corresponding character of the band. (b) Fermi surface plotted using XcrysDen visualization package \cite{XCrysden} (c). FS  map with $h\nu= 575$ eV. The superimposed dashed lines are the DFT FS sheets.
(d). FS map along cut 2 in (c), taken with $h\nu = 440 - 640$ eV in steps of 2.5 eV.  The superimposed dashed lines in (c) and (d) are
the FS from DFT calculations for $k_{||}<0$.   (e), (f). ARPES spectra along cut 1 in (c) at $h\nu= 575$ eV and $525$ eV, respectively. The cut at $h\nu=575$ eV ($h\nu=525$ eV) is close to the $Z$-$\Gamma$ ($\Gamma$-$M$) direction.
(g), (h). ARPES spectra along  cut 2 in (c) at $k_z = 21$ and  $k_z = 20$, respectively.  The DFT bands are renormalized by a factor $(W_{DFT}/W)_\text{ARPES}=1.5$.}\label{LaFe2P2_BS}
\end{figure*}
%

In this letter, we emphasize the conceptual difference between the
nominal valence count corresponding to the band filling and the
effective orbital occupancies of the correlated states not only by doping but also by variations of the hybridisations. Based on
a combined experimental and theoretical spectroscopic study, we show
that the latter is a more reliable tuning parameter for electronic 
correlations.
We perform SX-ARPES for a series of stoichiometric 122 pnictides, namely, 
LaFe$_2$P$_2$, CaFe$_2$P$_2$ and BaFe$_2$As$_2$. The use of  SX-ARPES, 
with its increased probing depth compared to ARPES performed with 
ultraviolet light as the excitation source (UV-ARPES),  is essential 
for the determination of the bulk electronic structure of these materials 
since it has been demonstrated that they show a surface state when 
measured in the UV range \cite{Heumen2011, Razzoli2012, Richard2014}. 
Moreover, we combine the SX-ARPES results with DFT+DMFT theoretical 
calculations to show how the electronic correlation effects are 
associated with the detailed electronic structure and 
filling in the studied series  of stoichiometric 122 pnictides.   
In particular, when $d$-orbital occupations are considered
rather than nominal chemical valences a clear link between the
evolution of the strength of electronic correlations and filling 
emerges. Indeed, contrarily to naive valence counting arguments, we 
find that the substitution of La with Ca does not change the $d$-orbital 
occupation, whereas the seemingly isovalent substitution of As by P 
does. Furthermore, we  identify the source of the latter change in the 
occupation, namely the change in bonding (BB) antibonding splitting of 
the pnictogen atoms due to the isovalent substitution.  Our results 
reveal that the   changes in the  correlation strength are mainly driven 
by unexpected changes in the orbital-resolved occupancy of the bands 
close to  $E_F$, suggesting that such changes may be fundamental in determining the superconducting properties of a given compound.

The SX-ARPES experiments were performed at the Advanced Resonant Spectroscopies (ADRESS) beamline at the Swiss Light Source (SLS). The experimental geometry is described in \cite{Strocov2014}. 
Data were collected using circularly-polarized light with an  overall energy resolution  of 50-80 meV. The samples were cleaved \textit{in situ} at 11 K  and measured in a vacuum always better than $5 \times 10^{-11}$ mbar. The \textit{\textbf{k}} values are determined by taking into account the photon momentum and are expressed in units of ($2\pi/a$, $2\pi/a$, $2\pi/c$). The $k_z$ values were extracted by using the free-electron final-state approximation with an inner potential $V_0=13$ for LaFe$_2$P$_2$ and CaFe$_2$P$_2$, and $V_0=14$ for BaFe$_2$As$_2$. All the intensity maps shown in the main text are obtained by integrating the ARPES spectral weight in an energy window of $E_F\pm 10$ meV.

Figure \ref{LaFe2P2_BS}(a) shows the band structure of  LaFe$_2$P$_2$ 
obtained using the Wien2k software package \cite{Wien2k}. 
The lattice constants used in the calculations are listed in Table \ref{Table_DFT}. The calculated Fermi surface (FS) [Fig. \ref{LaFe2P2_BS}(b)] is similar to that of LaRu$_2$P$_2$ presented in \cite{Razzoli2012} with the exception that there is an additional band ($\gamma$ in Fig. \ref{LaFe2P2_BS}) that crosses $E_F$ along the $\Gamma$-$X$ direction in LaFe$_2$P$_2$ and forms a small hole pocket at the zone corner.

In Figs.~\ref{LaFe2P2_BS}(c) and ~\ref{LaFe2P2_BS}(d) we plot the ARPES spectral weight mapping at $E_F$ near the ($k_x$, $k_y$, $21$) and in the ($k_{||}$, $2 - k_{||}$, $k_z$)  plane. The $k_z$ values were extracted by using the free-electron final-state approximation (FEFSA) \cite{Hufner}. The overlaid dashed lines show the FS from DFT calculations for $k_x<0$ (The FS for $k_x>0$ is obtained by reflection with respect to the $k_x=0$ axis). The $k_z$ variation as a function of ($k_x$, $k_y$) in the ARPES measurements with fixed photon energy has been taken into account according to the FEFSA. The observed FS  is in good agreement with DFT calculations and it is highly three-dimensional, similarly to the case of LaRu$_2$P$_2$ \cite{Razzoli2012}.

\begin{figure}[b]
\includegraphics[width=0.50\textwidth]{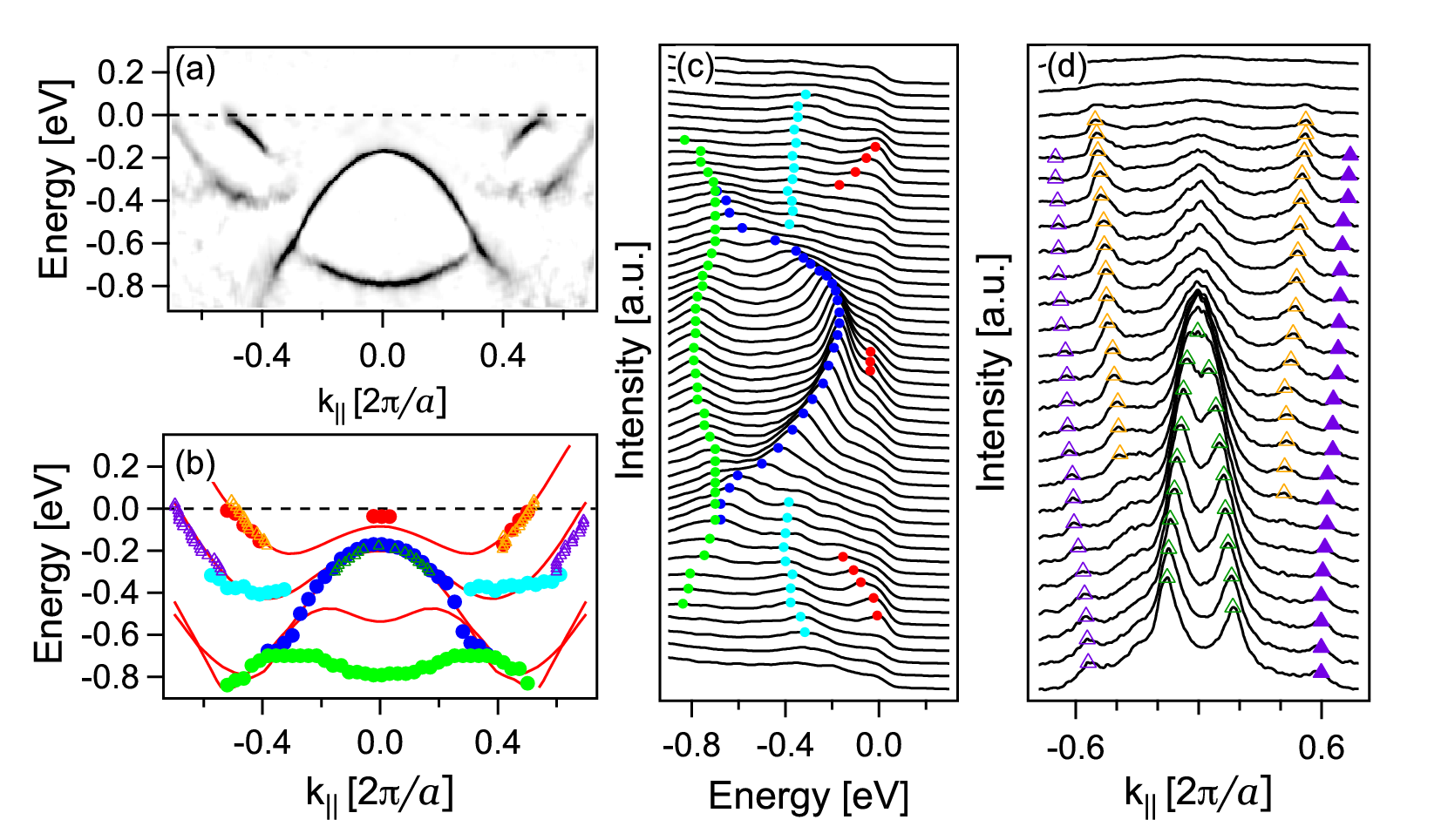}
\caption{Curvature, EDCs and MDCs analysis for the  $\Gamma - M$ direction. (a) Curvature intensity plot obtained from Fig. \ref{LaFe2P2_BS}(f). (b) MDCs and EDCs peaks supperimposed on the  renormalized DFT bands (c)-(d) EDCs and MDCs from Fig. \ref{LaFe2P2_BS}(f).}\label{LaFe2P2_CurvEdcsMdcs}
\end{figure}
%

\begin{figure*}[t]
\centering
\includegraphics[width=1\textwidth]{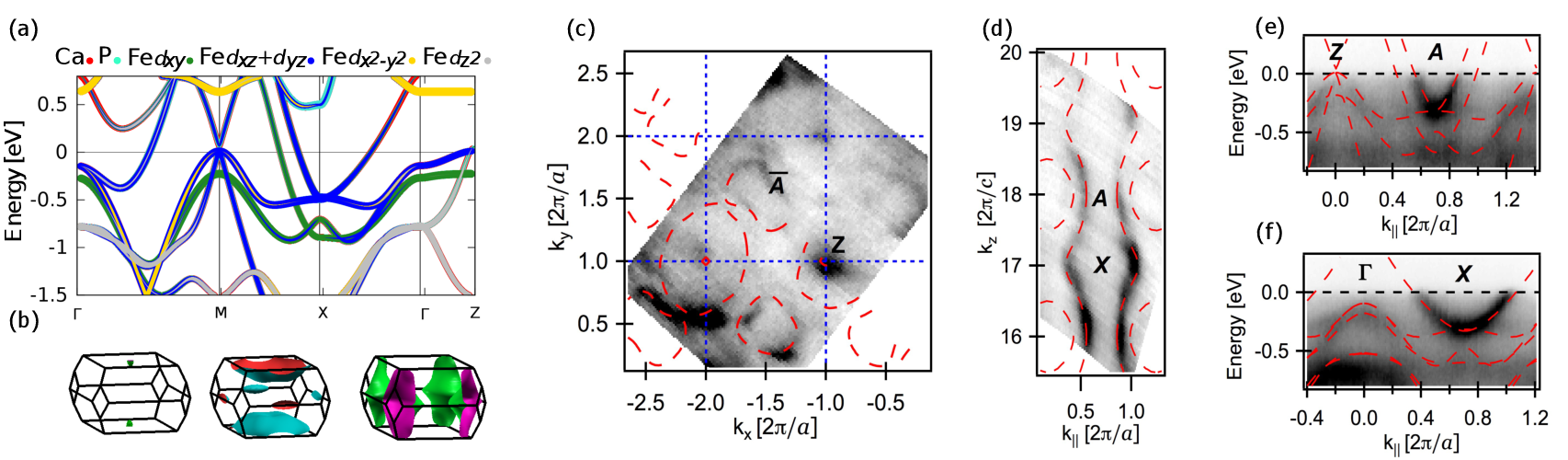}
\caption{DFT band structure and SX-ARPES intensity maps of CaFe$_2$P$_2$. (a) Band structure  along high symmetry lines.  (b) Three-dimensional FS sheets of different bands.   (c) FS  map with $h\nu= 525$ eV. (d) Intensity maps in the ($k_{||} - 2$, $1-k_{||}$, $k_z$) plane, taken with $h\nu = 440 - 640$ eV in step of 2.5 eV. 
(e), (f) Spectra along the ($k_{||} - 2$, $1-k_{||}$) direction with  $k_z=18$ (Z-A direction) and $k_z=17$  ($\Gamma$-X direction), respectively. The superimposed dashed lines are the DFT  bands  renormalized by a factor $\sim 1.5$.  }\label{CaFe2P2}
\end{figure*}

The  ARPES intensity as a function of energy and $k$ along cut 1 in Fig.~\ref{LaFe2P2_BS}(c) at $h\nu= 575$ eV ($k_z\sim 21$)  and $525$ eV ($k_z\sim 20$)  are shown in Figs.~\ref{LaFe2P2_BS}(e)-\ref{LaFe2P2_BS}(f), respectively. Since the value of $k_z$ changes substantially along cut 2 in Fig. \ref{LaFe2P2_BS}(c) ($\Delta k_z \sim 0.5$), to trace the band dispersion along the $\Gamma$-$X$ and $Z$-$A$ symmetry axes we made interpolations between spectra at different photon energies to obtain the spectra with constant $k_z$ shown in Fig. \ref{LaFe2P2_BS}(g)-\ref{LaFe2P2_BS}(h). The DFT band structure (dashed lines)  are superimposed on the data.  Contrary to the case of LaRu$_2$P$_2$, where DFT reproduces the measured band dispersion quite well, the overall agreement between the ARPES spectra and the calculated electronic structure of LaFe$_2$P$_2$ becomes reasonable only once the bands close to $E_F$ are renormalized by a factor $(W_{DFT}/W)_\text{ARPES}= 1.5\pm 0.1$ (red dashed lines). Assuming a purely local and orbital-independent many-body self-energy to be at the origin of this renormalisation, we estimate a quasiparticle residue of $Z_{ARPES}=0.67 \pm 0.05$ \cite{info1}. All the bands close to the $E_F$ are renormalized by this  same factor, as evident from the curvature \cite{Zhang2011}, EDCs and MDCs analysis of Fig. \ref{LaFe2P2_CurvEdcsMdcs}.

The decrease of the bare bandwidth, due to the change in the principal quantum number of the electrons close to the $E_F$, as well as the increase of  Hubbard interaction $U$ and Hund's coupling $J$ (see Table \ref{Table_Ren}) upon isovalent substitution of Ru with Fe  significantly contributes to the increasing  correlations in the system.
However, the observed band renormalization factor in LaFe$_2$P$_2$ is smaller than in the antiferromagnetic (AFM) compound (e.g. BaFe$_2$As$_2$) and in the superconducting 122 Fe-based pnictides (e.g. Ba$_{1-x}$K$_{x}$Fe$_2$As$_2$), indicating that the above changes induced by the Ru $ \rightarrow$ Fe substitution, are not the only mechanisms responsible for the increase of the strength of the electronic correlations observed in these systems.

{\renewcommand{\arraystretch}{2}
\begin{table}[b]
\caption{The lattice constants and internal atomic positions for LaRu$_2$P$_2$ \cite{Jeitschko1987}, LaFe$_2$P$_2$ \cite{Jeitschko1985}, CaFe$_2$P$_2$ \cite{Mewis1980}, and BaFe$_2$As$_2$ \cite{Huang2008}. }
\label{Table_DFT} 
\begin{ruledtabular}
\begin{tabular}{|c c c c|}
Sample  & $a ($\AA$)$ & $c ($\AA$)$ & $z_\text{Pn}$ \\
\hline
LaRu$_2$P$_2$   &  $4.031$ & $10.675$ &  $0.3593$\\
\hline
 LaFe$_2$P$_2$   &  $3.841$ & $10.982$ &  $0.3554$\\
\hline
 CaFe$_2$P$_2$   & $3.855$ & $9.985$  & $0.3643$\\
\hline
BaFe$_2$As$_2$   & $3.957$ & $12.969$ & $0.354$\\
\end{tabular}
\end{ruledtabular}
\end{table}}

To investigate the effect of hole doping in the system, we performed a similar study for CaFe$_2$P$_2$, which has one more hole per unit cell compared to LaFe$_2$P$_2$ and LaRu$_2$P$_2$. Similar to these compounds, the 3D FS of CaFe$_2$P$_2$ is in good agreement with the DFT calculations [see Figs. \ref{CaFe2P2}(a)-\ref{CaFe2P2}(d)], while the band structure has to be renormalized by a factor of $(W_{DFT}/W)_\text{ARPES}=1.5 \pm 0.1$ [Figs.\ref{CaFe2P2}(e)-(f)]. 
This result is quite surprising, since the bandwidth renormalization is not influenced by the formal decrease in the number of the electrons due to the La to Ca substitution, contrary to the expected behavior of a system moving toward half filling \cite{Ishida2010}. However this is explained once the multiband nature of these systems is taken into account. Comparison between Figs. \ref{LaFe2P2_BS}(a) and  \ref{CaFe2P2}(a) shows that the substitution of La with Ca does not result in a rigid shift of all the bands. Rather, only one band [$\gamma$ in Figure \ref{LaFe2P2_BS}(a)] is fully pushed above  $E_F$, while the others (with Fe $d$ character) crossing  $E_F$ are qualitatively unchanged.
This is mainly due to the fact that the $\gamma$ band is very sensitive to the La to Ca substitution since it is essentially a La (Ca) band hybridizing (strongly in the case of La) with Fe $d$ and P $p$ states. The filling of the $d$ bands is thus barely changed even if the total number of electron in the unit cell is decreased by 1.
We note here that the comparison between  our finding  and the mass enhancement measured by various  quantum oscillation experiments  ($m/m_{DFT}$ in Table \ref{Table_Ren}) shows that while the electron-phonon interactions are responsible for the main contribution to the mass enhancement in LaRu$_2$P$_2$, most likely causing its low-$T_c$ superconducting state, they are negligible in LaFe$_2$P$_2$ and CaFe$_2$P$_2$, which are indeed not superconducting.

{\renewcommand{\arraystretch}{2}
\begin{table*}
\centering
\caption{Critical temperature, ground state, bandwidth renormalisation with respect to DFT extracted from ARPES, mass enhancements reported in the literature, Hubbard $U$ and Hund's coupling $J$ calculated from cRPA,  bare bandwidth obtained from DFT , vertical pnictogen-pnictogen distance and nominal occupation of 3$d$-shell for the samples LaRu$_2$P$_2$, LaFe$_2$P$_2$, CaFe$_2$P$_2$, and BaFe$_2$As$_2$.}
\begin{ruledtabular}
\label{Table_Ren} 
\begin{tabular}{|c c c c c c c c c c|}
sample & $T_c$ (K) & \specialcell{Ground\\*[-0.2 cm] state}	& $\frac{\text{W}_{DFT}}{\text{W}_{ARPES}}$ &   $\frac{m}{m_{DFT}}$ &  $U_{cRPA}$ (eV)  & $J_{cRPA}$ (eV) & W$_{DFT}$ (eV)  & $z^\perp_{\text{Pn-Pn}}$ (\AA) & $n_d^{nom}$ \\
\hline
 \quad LaRu$_2$P$_2$ & 4 {\cite{Jeitschko1985}}  &  low $T_c$ SC & 1 {\cite{Razzoli2012}}  &  2 {\cite{Moll2012}}  & 1.89 & 0.502 & 9.4 & 3.00 {\cite{Jeitschko1985}}  & 6.5\\
\hline
 \quad LaFe$_2$P$_2$ &     & non-SC & 1.5 & 1.8 {\cite{Moll2012}}  & 2.035 & 0.695 & 7.55 & 3.20 {\cite{Jeitschko1985}}  & 6.5\\
\hline
 \quad CaFe$_2$P$_2$ &  & non-SC & 1.5 & 1.5 {\cite{Coldea2009}}  & 2.632 & 0.723 & 7.50 & 2.71 {\cite{Mewis1980}}  & 6\\
\hline
\quad BaFe$_2$As$_2$ & 142 {\cite{Huang2008}}  & AFM & 2.1 &  2-3 {\cite{Terashima2011}}  & 2.572  & 0.78 & 7.4 & 3.78 {\cite{Huang2008}}  & 6\\
\end{tabular}
\end{ruledtabular}
\end{table*}}

\begin{figure}[t]
\centering
\includegraphics[width=0.5\textwidth]{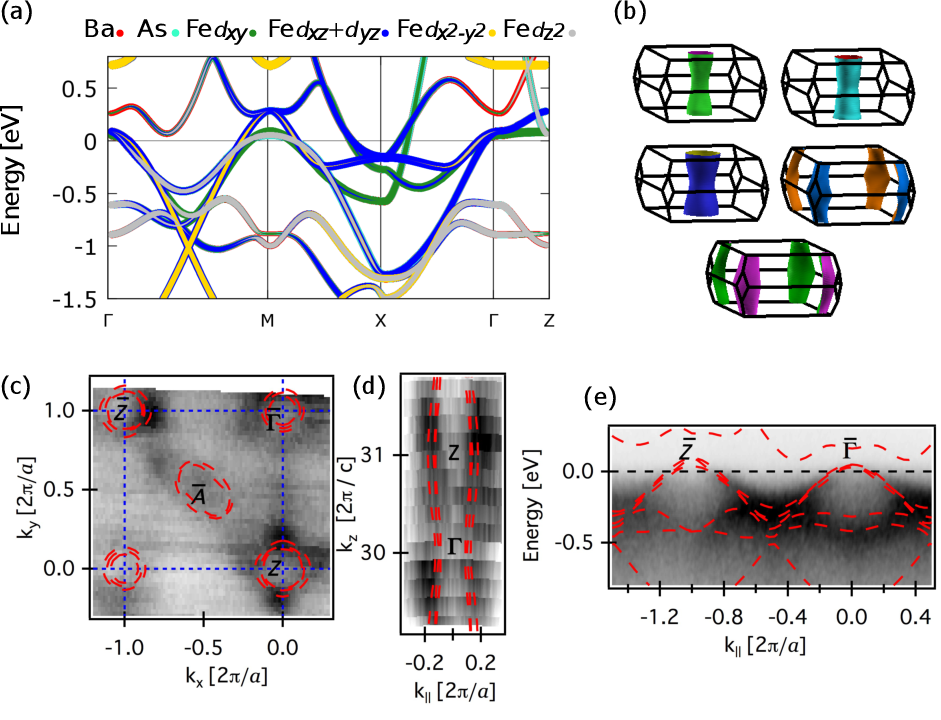}
\caption{DFT band structure and SX-ARPES intensity maps of BaFe$_2$As$_2$. (a) Band structure  along high symmetry lines. (b) Three-dimensional FS sheets of different bands. (c) FS  map of BaFe$_2$As$_2$ with $h\nu= 840$ eV.  (d)  Intensity maps in the ($k_{||}$, $0$, $k_z$) plane, taken with $h\nu = 720 - 930$ eV in step of $10$ eV. 
(e) Spectrum along the ($k_{||}$, 1) direction with at $h\nu=840$ eV ($k_z\sim 30$). The superimposed dashed lines are the DFT calculations 
 renormalized by a factor $\sim 2.1$.}\label{BaFe2As2}
\end{figure}

\begin{figure}[t]
\includegraphics[width=0.49\textwidth]{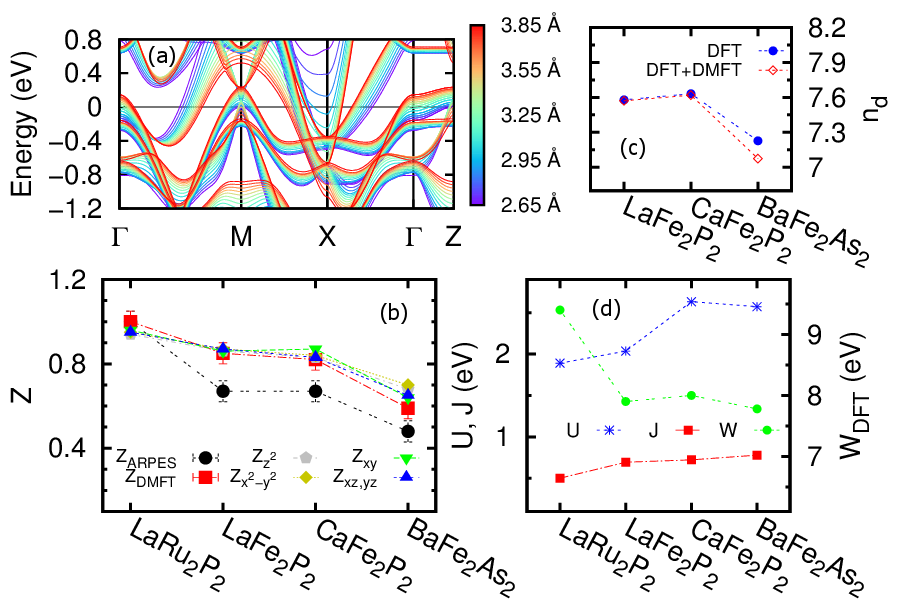}
\caption{DFT and DFT+DMFT calculation results. 
(a) DFT calculation in CaFe$_2$P$_2$ as a function of the P-P distance.
(b) Experimental and theoretical values of  $Z$. (c) Orbital dependent occupation of Wannier $d$ orbitals within DFT and DFT+DMFT.  (d) Calculated values of $U_{cRPA}$, $J_{cRPA}$ and $W_{DFT}$.}\label{DMFT}
\end{figure}

Having discussed the role of the La to Ca substitution, it is now important  to clarify the origin of the change in the bandwidth renormalization from CaFe$_2$P$_2$ to BaFe$_2$As$_2$. The nominal filling, the screened Coulomb interaction $U$ and the Hund's coupling $J$ [see Table \ref{Table_Ren}] are very similar in the two systems, suggesting similarities in the electronic correlations. However our measurements for BaFe$_2$As$_2$ [see Fig. \ref{BaFe2As2}], 
for which a similar analysis as above leads to identifying  a bandwidth renormalization $\sim 2.1 \pm 0.2$ ($Z_{ARPES}=0.48 \pm 0.05$) in agreement with the existent ARPES measurements in the UV range \cite{Richard2011}, indicate that there is a sizable increase in the bandwidth renormalization. 
Again the change in the bandwidth renormalization can be understood once the effective  filling of the $d$ bands is considered, instead of the nominal value. A first indication that the actual filling of the $d$ bands is decreased in BaFe$_2$As$_2$ comes from comparing the DFT calculations. Focusing  on the $d_{xz}$ ($d_{yz}$) bands, the comparison between Figs.\ref{CaFe2P2}(a) and \ref{BaFe2As2}(a) reveals that the bands at -250 meV in CaFe$_2$P$_2$  are very similar to the ones at E$_F$ in BaFe$_2$As$_2$, \textit{i.e.,} these bands appear shifted by 250 meV with respect to each others. The change in the  band structure in this case is mainly due to the P to As substitution (similarly to the case of CaFe$_2$P$_2$ and CaFe$_2$As$_2$ \cite{Coldea2009}), which results in a change in the distance between the pnictide atoms Pn (Pn = P, As) of two adjacent Fe-Pn layers. In  case of smaller distance (in CaFe$_2$P$_2$) the stronger bonding (BB) antibonding (AB) splitting of the Pn bands results in the AB bands being above $E_F$ [in particular the band at 0.5 eV at the X point in Fig. \ref{CaFe2P2}(a) \cite{Andersen2011}. In  case of larger distance, as in  BaFe$_2$As$_2$, the splitting is smaller and the same band now crosses $E_F$. To compensate the increase of the carriers in this band, the other bands shift of towards $E_F$.  A clear demonstration of this mechanism is shown in Fig. \ref{DMFT}(a), where the DFT calculations for CaFe$_2$P$_2$ are performed as a function of the P-P distance for a fixed  P-Fe distance.

To summarize, we have shown that while the weak strength of correlations in LaRu$_2$P$_2$ is mainly due to its larger bare bandwidth W$_{DFT}$ [see Table \ref{Table_Ren} and  Fig. \ref{DMFT}(d)] the evolution of correlations strength in LaFe$_2$P$_2$, CaFe$_2$P$_2$ and BaFe$_2$As$_2$  is driven by the effective filling of the $d$ orbitals  crossing the $E_F$.
The strength of the correlations remains constant in LaFe$_2$P$_2$ and CaFe$_2$P$_2$ due to the similar filling, and increases in BaFe$_2$As$_2$ when the filling decreases [see Fig. \ref{DMFT}(b)]. This result differs from what would be naively expected  from the nominal Fe $3d$ band filling, $i.e.$, the same strength of correlations in the $3d^{6.5}$ LaRu$_2$P$_2$ and LaFe$_2$P$_2$ and bigger $(W_{DFT}/W)_\text{ARPES}$ in CaFe$_2$P$_2$ and BaFe$_2$As$_2$, which are both nominally in the $3d^6$ configuration. 
We underline that the observed occupation-driven trend in the correlations confirms the link between cuprate and pnictides phase diagrams. The average orbital doping from our DFT calculations are 0.29 and 0.28 for LaFe$_2$P$_2$ and CaFe$_2$P$_2$, respectively.  These results would put them in the overdoped non-superconducting side of the unified phase diagram proposed in \cite{DeMedici2014}, where the correlations are not strong enough to allow the development of unconventional superconductivity and to display differentiation of bandwidth renormalization i.e., selective Mottness, for electrons with different band characters.

The qualitative description above can be confirmed by theoretical calculations using  the combined  DFT+DMFT method which gives access to the spectral properties of these compounds.  The DMFT calculations were performed in the implementation of \cite{Aichhorn2009}, using the the  hybridization expansion continuous-time quantum Monte Carlo algorithm \cite{Werner2006} as implemented in the TRIQS toolkit \cite{TRIQS}. 
Localized Wannier orbitals were built by truncating the  expansion of the initial atomic-like orbitals to an energy window $\mathcal{W}$, chosen here as $\mathcal{W}=[-7.5,3]$ eV (see \cite{Aichhorn2009} for details). Calculations have been carried out at an inverse temperature  of $\beta=40$ eV$^{-1}$, in the paramagnetic phases of the compounds.  The calculations of the Hubbard interactions and Hund's coupling (or more generally the Slater integrals F$_0$, F$_2$, F$_4$) have been performed using  the  constrained random phase approximation (cRPA) method in the implementation of \cite{Vaugier2012}.

\begin{figure*}[t]
\includegraphics[width=1\textwidth]{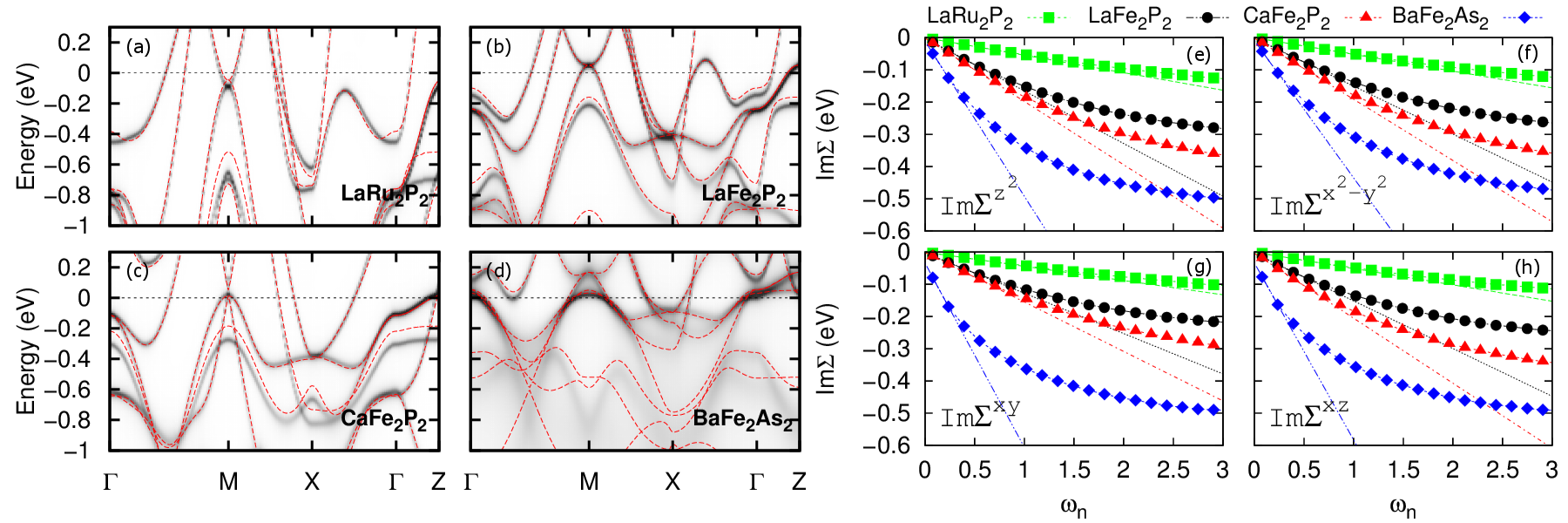}
\caption{DFT+DMFT spectral functions and self energies.
(a)-(d) Spectral functions calculated with the DFT+DMFT method along high symmetry lines for LaRu$_{2}$P$_{2}$, LaFe$_{2}$P$_{2}$, CaFe$_{2}$P$_{2}$ and BaFe$_{2}$As$_{2}$, respectively. The DFT calculations superimposed are renormalized by the factors $(Z_{\text{DMFT}})^{-1}$, with $Z_{\text{DMFT}}$ displayed in Fig. \ref{DMFT} (b). (e)-(h) Imaginary parts of the local many-body
self-energy on the imaginary frequency axis. }\label{DMFT_Spectral}
\end{figure*}

The values obtained for the interactions are shown in Fig. \ref{DMFT}(b). The DMFT  spectral functions  for the various samples are presented in Fig. \ref{DMFT_Spectral} (a)-(d). The DFT band structures, overlaid on the respective spectral functions are renormalized by a factor ($W_{DFT}/W$)$_{\text{DMFT}}$.  The resulting quasiparticle residues (denoted $Z_{DMFT}$) agree well with the $Z_{m}$ values ($m=z^2, x^2-y^2, xy, xz, yz$) directly extracted from the self-energy on the Matsubara grid as $Z_m= \big\{ {1-\text{Im}[\frac{d\Sigma_m(i\omega_n)}{d\omega_n}|_{\omega_n\rightarrow 0}]} \big\}^{-1}$ [see Fig. \ref{DMFT}(b)].
The calculated behavior agrees well with the experimental one; the correlations are negligible in LaRu$_2$P$_2$, increase in LaFe$_2$P$_2$, remain constant in CaFe$_2$P$_2$, and reach a maximum in BaFe$_2$As$_2$. Fig. \ref{DMFT_Spectral} (e)-(h) shows the imaginary parts of the local many-body
self-energy on the imaginary frequency axis, a quantity which in the
Fermi liquid regime exhibits linear behavior at low energies (and a 
slope related to the quasiparticle residues as above). These curves
show that the DMFT calculations display the trend of correlation
strength for our series of compounds in two ways:
not only do the self-energies increase in overall value, with the steeper slopes for BaFe$_2$As$_2$ corresponding to the stronger quasi-particle renormalisations (smaller $Z_m$), but also does the energetic {\it extent} of the linear regime decrease. In BaFe$_2$As$_2$, coherent quasi-particles exist only on very low energy scales, while the other compounds display larger coherence scales. In the spectral functions, the signature of  this effect is the overall quite substantial broadening observed in BaFe$_2$As$_2$, as compared to the other compounds.
The small underestimation of  $(W_{DFT}/W)$ in the DMFT calculations as compared to experiment might be due to the local nature of the DMFT method \cite{Georges1996} or the absence of dynamical screening in the calculations \cite{Werner2012, Casula2012, vanRoekeghem2014}. 
We also note that there is a sizable change in $U_{cRPA}$ and $J_{cRPA}$ only upon the Ru to Fe substitution and that, even in this case, it does not have a large influence on the correlations in our calculations. Indeed the calculated $Z_{m}$ for the LaRu$_2$P$_2$  are slightly changed by an increase in $U_{cRPA}$ and $J_{cRPA}$ (values obtained using the same $U_{cRPA}$ and $J_{cRPA}$  used for BaFe$_2$As$_2$ are  $Z_{z^2}=0.90, Z_{x^2-y^2}=0.91, Z_{xy}=0.92, Z_{xz, yz}=0.90$). 
The evolution of the filling of the Wannier  $d$ orbitals, both before and after the DFT-DMFT calculations, is shown in Fig. \ref{DMFT}(c). The filling of the orbitals is constant in LaFe$_2$P$_2$ and  CaFe$_2$P$_2$ and decreases in BaFe$_2$As$_2$, in agreement with the qualitative conclusions drawn from the band structure calculations.

We finally remark that a previous work has proposed CaFe$_2$P$_2$ as structural analogue of the collapsed tetragonal (CT) non-magnetic phase of CaFe$_2$As$_2$ \cite{Coldea2009}.  However the measured  correlation strength observed here for CaFe$_2$P$_2$ is smaller than the value reported for the CT phase of CaFe$_2$As$_2$ \cite{Gofryk2014, Mandal2014, Diehl2014}, indicating that there might be a fundamental difference between the two systems. 

In conclusion, we showed that the evolution of the electronic correlations in the presented series of  pnictides differs from the expected behavior inferred from the nominal filling of the correlated orbitals. The experimental trend can be qualitatively understood  based on the effective filling of the correlated orbitals and quantitatively reproduced by  our DMFT calculations.  We also demonstrated that the observed occupation driven trend in the correlations supports the recently proposed link between the unconventional superconductivity in the cuprates and Fe-pnictides.
Finally, the high sensitivity of the correlations to small changes in the  filling opens new ways of tuning the strength of electronic correlations in transition metal pnictide systems by systematic chemical substitutions, pressure or constrains.

\begin{acknowledgments}
This project was supported by the Swiss National Science Foundation (grant no. 200021-137783),  through MaNEP (grant no. 200020-105151) and Division II, the French ANR under project PNICTIDES, IDRIS/GENCI under project 1393, the European Research Council under project number 617196
and the Cai Yuanpei program.
P. R. and H. D. acknowledge grants from MOST (Nos. 2010CB923000, 2011CBA001000, 2011CB921701, 2013CB921700 and 2012CB821403) and NSFC (Nos. 11004232, 11234014 and 11274362) from China.
The DFT+DMFT were performed at DALCO cluster of the Univerity of Fribourg.  We thank Pawel Bednarek for the support.
\end{acknowledgments}



\begin{thebibliography}{1}
\bibitem{info1} The bandwidth renomalization factor is symbolically indicated as $(W_{DFT}/W)_{i}$, with the index i ($i=$ARPES, DMFT) specifying whether it is obtained by re-scaling the DFT bands to match the experimental spectral function or the calculated one. The associated  quasiparticle residue  $Z$ is defined  as $Z_i \equiv (W_{DFT}/W)_i^{-1}$.
\end{thebibliography}
\end{document}